\documentstyle[12pt,psfig]{article}
\title{\bf A Search for Chemical Signatures of Galactic Mergers}
\author{Inese I. Ivans $^1$\thanks{iivans@astro.as.utexas.edu}, 
	Bruce Carney $^2$, Luisa de Almeida $^2$,
	and Chris Sneden $^1$\\
\vspace{1cm}\\
\normalsize $^1$Department of Astronomy and McDonald Observatory, 
	University of Texas at Austin, USA\\
\normalsize $^2$Department of Physics and Astronomy, 
	University of North Carolina at Chapel Hill, USA\\}
\date{\mbox{}}
\begin{document}
\maketitle
\pagestyle{empty}
\def\bull{\vrule height .9ex width .8ex depth -.1ex}
\makeatletter
\def\ps@plain{\let\@mkboth\gobbletwo
\def\@oddhead{}\def\@oddfoot{\hfil\tiny\bull\quad
``The Galactic Halo: From Globular Clusters to Field Stars'';
35$^{\mbox{\rm th}}$ Li\`ege\ Int.\ Astroph.\ Coll., 1999\quad\bull}%
\def\@evenhead{}\let\@evenfoot\@oddfoot}
\makeatother
\def\beginrefer{\section*{References}%
\begin{quotation}\mbox{}\par}
\def\refer#1\par{{\setlength{\parindent}{-\leftmargin}\indent#1\par}}
\def\endrefer{\end{quotation}}
{\noindent\small{\bf Abstract:} 
We present preliminary chemical abundance analyses of a group of 
high-velocity metal-poor stars.  This report describes our work in progress.
We examine the derived abundances in the context of previous studies of 
metal-poor stars with unusual abundances and/or extreme galactocentric 
orbits.  Included in this sample are BD+80~245, a star previously known to 
have unusually low $\alpha$-element abundances, and G4-36, a new low-$\alpha$ 
star.}
\section{Overview}

A number of very high-velocity metal-poor field stars have been discovered 
that have very unusual ratios of alpha-elements to iron.  The stars
discovered to date all have large apogalacticon distances, and so the 
unusual abundance ratios may suggest a chemical ``signature'' of previous 
merger or accretion events.  That is, these stars may have originated within 
a satellite galaxy or galaxies that experienced a different nucleosynthetic 
chemical evolution history than the Milky Way and which were later accreted 
by it.  

We have observed a sample of over two dozen high-velocity metal-poor field 
stars using high resolution echelle spectrographs at the KPNO, CTIO, and 
McDonald Observatories.  The stars were selected from Ryan \& Norris (1991), 
Carney {\it et al} (1994), as well as unpublished studies yielding private 
catalogs of metal-poor stars.  So far, we have analysed the more metal-poor 
stars and, for those stars in common with previous studies, we obtain 
results that agree well with those in the literature.  Included in our 
study are a re-analysis of BD+80~245, a star previously known to have low 
$\alpha$-element ratios (Carney {\it et al} 1997) and G4-36, a new 
low-$\alpha$ star discovered by James (1998).

\section{BD+80~245: re-analysis of the ``discovery'' star}

Carney {\it et al} (1997) reported the discovery of an $\alpha$-element
poor star as a result of their search for low-metallicity disk stars.  We 
have since aquired new spectra of the star, with a resolution of 60,000
and a SNR of $\sim$200.  We have independently re-derived the abundances, 
using a different linelist, and largely confirm the previous analysis, as 
well as expand the abundance list to include additional key elements.

\begin{small}
\begin{table}
\begin{center}
\begin{tabular}{|c|c|c|c|c|c|c|c|} \hline
Value & Teff & log~g & v$_{micro}$ & $\left[{\rm FeI/H}\right]$ & $\left[{\rm FeII/H}\right]$ & n(Li) \\ \hline
Old   & 5400 & 3.20  & 1.50        & --1.86                     & --1.96                      & 1.75 \\
New   & 5425 & 3.25  & 1.35        & --1.85                     & --1.86                      & 1.36 \\ \hline
\end{tabular}
\end{center}
\caption{A Comparison of Previous and New Values Derived for BD+80~245}
\begin{center}
\begin{tabular}{|c|c|c|c|c|c|c|c|c|c|c|c|c|c|c|c|c|c|} \hline
      &     O    &    Na    &    Mg  &    Al    &    Si    &   Ca   &     Sc   &   Ti   &    Cr    &    Mn    &    Ni    &   Ba   \\ \hline
Old   & +0.39    & $\cdots$ & --0.31 & $\cdots$ & $\cdots$ & --0.30 & $\cdots$ & --0.26 & $\cdots$ & $\cdots$ & $\cdots$ & --1.84 \\ 
New   & $\cdots$ & --0.49   & --0.21 & --1.31   & --0.07   & --0.22 & --0.26   & --0.26 & --0.13   & --0.13   & --0.10   & --1.68 \\ \hline 
\end{tabular}
\end{center}
\caption{Previous and New [el/Fe] Results for BD+80~245}
\end{table}
\end{small}

\section{G4-36: a star of low Na, Mg, Ca, but high Ni}

Among our observations, we included a star of unusual abundances discovered 
by James (1998).  In addition to confirming the unusually low Na, Mg, Ca and 
Ba abundances determined by James (PhD Thesis in preparation, Univ. of 
Texas), we find significantly enhanced Ni.  This is contrary to the results 
of Nissen \& Schuster who found that the abundances of Ni to be strongly 
correlated with Na in their low-$\alpha$ stars.

\begin{figure}[bhpt]
\centerline{\psfig{file=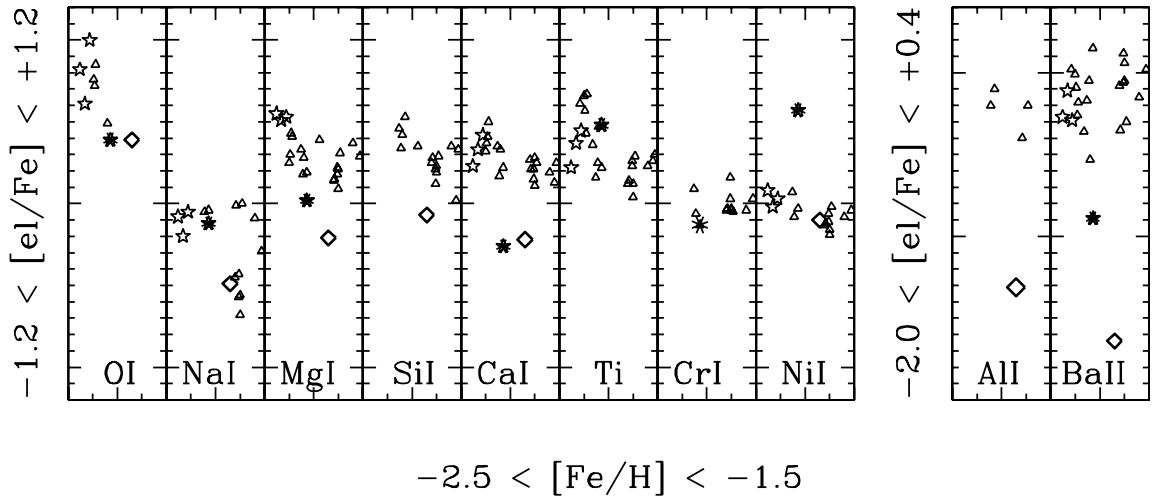,clip=}}
\caption{The mean abundances [el/Fe], relative to the solar values, of a 
number of elements in stars with metallicities similar to those of G4-36 
and BD+80~245.  The abscissa for each section covers the range --2.5 $<$ 
[Fe/H] $<$ --1.5.  Note that the plots on the left and the right are scaled 
to illustrate the same relative ranges in abundance.  In order to provide a 
context for some of the unusual abundances we derived in this study for G4-36 
($*$) and BD+80~245 ($\Diamond$), we show the results of other stars 
($\star$) of this study, as well as those from Gratton \& Sneden (1988), 
Magain (1989), Carney {\it et al} (1997), Stephens (1999), and other stars of 
Rapo $>$ 15kpc, as compiled by Carney {\it et al} (1997) ($\triangle$).}
\end{figure}

\section{Na vs. Ni}

As previously found by Nissen \& Schuster (1997), for most of the stars, 
there appears to be a good correlation between Na and Ni abundances with 
respect to iron.  As seen in figure 2, this correlation is also found in the 
Stephens (1999) data.  However, a few exceptions stand out: the low globular 
cluster abundances found by Brown {\it et al} (1997) for Rup~106 and Pal~12; 
the unusual star BD+80~245 found by Carney {\it et al} (1997); as well as 
BD+24 1676, a star which seems to show mild enhancements in Mg, Ca, and 
Ti with respect to its iron abundance.  Including G4-36, yet another star 
of interesting abundance ratios, these ``unusual'' stars seem to fit 
together, and correspond to a Ni/Na correlation as well, albeit a very 
different one from the rest of the stars.  Whether this is a real 
correlation, or simply a coincidence imposed by the small number of stars 
analysed so far, will be determined as the remainder of our two-dozen high 
velocity star sample is analysed.  However, if the secondary correlation 
exists, a Ni/Na discriminant such as this could be extremely useful as a 
proxy indicator to target stars of other unusual abundance ratios, thus 
alerting us to their different nucleosynthetic histories.

\begin{figure}[bhpt]
\centerline{\psfig{file=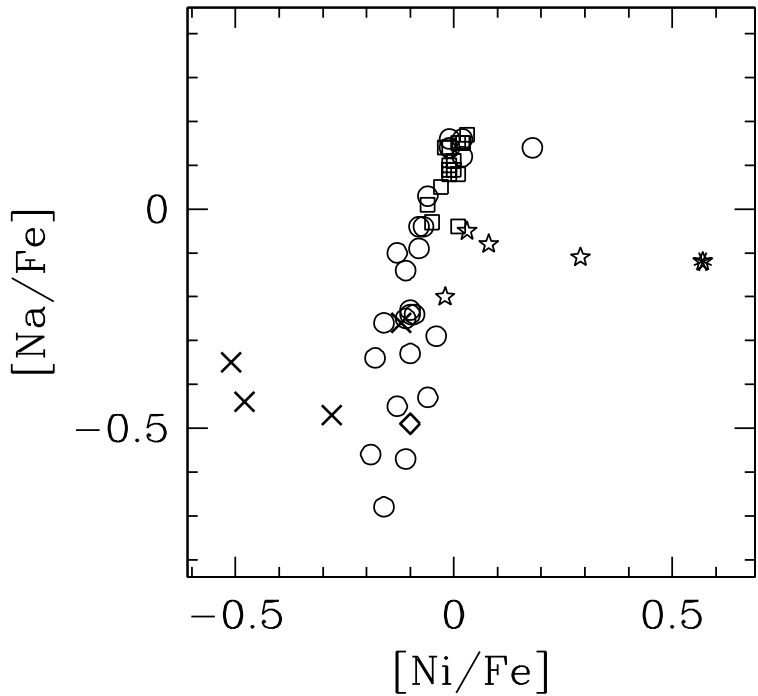,clip=}}
\caption{[Na/Fe] vs. [Ni/Fe].  We plot the results of G4-36 ($*$) and
BD+80~245 ($\Diamond$) and the other stars of this study ($\star$) in
the context of the halo stars ($\bigcirc$) and disk stars ($\Box$) 
studied by Carney {\it et al} (1997), Nissen \& Schuster (1997), and 
Stephens (1999), as well as the globular cluster stars ($\times$) 
studied by Brown {\it et al} (1997).  The overall correlation was 
observed by Nissen \& Schuster (1997) and the results of Stephens (1999) 
fit this well.  However, there are some points belonging to stars of 
unusual abundance ratios, that do not fit the main trend but, rather, 
could be fit by another slope.}
\end{figure}

\section{Possible Correlations with Kinematics}

The abundances of $\alpha$- and iron-peak elements as a function of the 
iron abundance do show abundance trends but, these trends are for the halo
stars only.  Most of the disk star abundances are constant across the 
almost 3 dex range in metallicity.  While no abundance trends are observed 
as a function of eccentricity, there does appear to be a larger scatter in 
the abundances with increasing eccentricity.  This is consistent with the 
relationship of the $\alpha$- and iron-peak elemental abundances with the
iron abundance: stars defined as members of the disk population have less 
scatter in the derived abundances.  A quantitative estimate of the scatter 
as a function of the kinematics will be derived once the abundance 
determinations have been made for our complete sample (for instance, our
initial analyses suggest that the scatter in the abundances of $\alpha$- 
and iron-peak elements as a function of the predicted perigalactocentric 
distances are larger for stars that get closer to the centre of the galaxy).  
Upon completion of the abundance analyses for the entire sample, we will
also investigate whether the unusual [$\alpha$/Fe] ratios have been found 
in a large enough sample of stars to identify the kinematics of the 
progenitor galaxy or galaxies. 
\section*{Acknowledgements}
We are indebted to Renee James notifying us of the unusual nature of G4-36 
prior to publication as well as supplying comparison abundance information.  
Ken Freeman has our appreciation for a helpful discussion and thoughtful 
suggestions regarding the analysis of the abundance trends and Sofia
Feltzing for her sharp eyes in spotting a misplaced point on our poster
display.  We also thank Daryl Wilmarth for his excellent technical help at 
the KPNO 4-m telescope as well as the subsequent assistance he provided in 
using the new and improved Thorium-Argon Spectral Atlas 
({\sf http://www.noao.edu/kpno/tharatlas/}).
 
\beginrefer
\refer Brown, Wallerstein, \& Vanture 1997, AJ 114, 180

\refer Carney {\it et al} 1994, AJ 107, 2240

\refer Carney {\it et al} 1997, AJ 114, 363

\refer Gratton \& Sneden 1988, A\&A 204, 193

\refer James 1998, BAAS 30, 1321

\refer James 1999, PhD Thesis in preparation, Univ. of Texas

\refer Magain 1989, A\&A 209, 211

\refer Nissen \& Schuster 1997, A\&A 326, 751

\refer Ryan \& Norris 1991, AJ 101, 1835

\refer Stephens 1999, AJ 117, 1771

\endrefer           
\end{document}